\documentclass[11pt,english]{article}
\usepackage[T1]{fontenc}
\usepackage[english]{babel}

\usepackage[top=1in,bottom=1in,left=0.75in,right=0.75in]{geometry}
\usepackage{times}
\usepackage{amsmath,amssymb,amsfonts}
\usepackage{graphicx}
\usepackage[super]{nth}
\usepackage{textcomp}

\usepackage[authoryear,round]{natbib}
\usepackage{booktabs}
\usepackage{caption}
\usepackage{subcaption}
\usepackage{float}
\usepackage{stfloats}

\usepackage{hyperref}
\hypersetup{
  colorlinks   = true,
  linkcolor    = blue,
  citecolor    = blue,
  urlcolor     = cyan
}

\usepackage{titlesec}
\titleformat{\section}
  {\normalfont\large\bfseries}{\thesection.}{1em}{}
\titleformat{\subsection}
  {\normalfont\normalsize\bfseries}{\thesubsection}{1em}{}
  \titleformat*{\subsubsection}{\normalfont}
\titlespacing*{\section}{0pt}{12pt}{6pt}
\titlespacing*{\subsection}{0pt}{12pt}{6pt}
\titlespacing*{\subsubsection}{0pt}{5pt}{4pt}

\usepackage[bottom,flushmargin]{footmisc}
\renewcommand{\thefootnote}{\arabic{footnote}}

\usepackage[stretch=10,shrink=10,step=1]{microtype}
\usepackage{enumitem}

\usepackage{tabularx}
\newcolumntype{L}{>{\raggedright\arraybackslash}X}
\begin{document}
\flushbottom
\twocolumn[
  \begin{center}
    {\LARGE The Science and Mythology of Sports Rankings:\\ Evaluating the methods, assumptions, and attributes used to \\rank NCAA teams\par}
    \vskip 1.5em
    {\large
      \lineskip 2em
      \begin{tabular}[t]{c}
        Lawrence Feldman\textsuperscript{\textdagger} \\
        \texttt{laf@laxnumbers.com}
      \end{tabular}\par}
      \vskip 1.5em
    Contributors:\\\textit{The Men's Division I Lacrosse Selection Criteria and Ranking Committee (SCR),\\ formerly chaired by Lars Tiffany.\textsuperscript{\textdaggerdbl}}
    \vskip 1.5em
  \end{center}
  
  \vskip 0.5em
  \noindent\textbf{Abstract:} Selecting a sports-ranking system is not a purely mathematical decision. It also requires agreement about what the ranking should mean and what institutional purpose it should serve. This paper distinguishes four related but different tasks in sports ranking: prediction, retrospective evaluation, championship selection, and seeding; compares the governing philosophies of major ranking families; and identifies recurring errors in ranking debates. A descriptive comparison of nine NCAA Division I sport settings illustrates how schedule density, scoring environment, cross-conference connectivity, head-to-head meetings, and common-opponent availability vary from sport to sport. The central premise is that no ranking methodology is value-free. Every system reflects choices about evidence, competitive merit, and which tradeoffs are acceptable. The paper therefore offers a structured, multi-objective framework for evaluating ranking systems according to their intended purpose/goal rather than naming one universally superior method. Its principal contribution is to make the philosophical and institutional assumptions behind ranking decisions explicit, comparable, and testable.
  
  \vskip 1.5em
]

\begingroup
\renewcommand{\thefootnote}{\fnsymbol{footnote}}

\footnotetext[2]{Lawrence Feldman served on the Coaches Committee for Division I men's lacrosse for the past three years as an advisor. He founded Laxpower, a web site that rated lacrosse teams from 1997 to 2019 and played lacrosse at the University of Pennsylvania. Before that, he earned his PhD in aerospace engineering and worked as an engineer/computer analyst in the USAF, Lawrence Livermore Laboratory (LLL), and Los Alamos National Laboratory (LANL).}
\footnotetext[3]{Lars Tiffany was the former head lacrosse coach at the University of Virginia, leading the team to national championships in ’19 and ‘21. He attended Brown University where he was team captain and later served as head coach.}

\endgroup

\setcounter{footnote}{0}

\section{Introduction: The purpose of rankings}
Sports rankings do more than simply place teams in order. They may, for instance, determine access to tournaments, influence seeding and scheduling, shape public perception, and limit access to institutional opportunities. However, debates about ranking systems often begin with disagreement about results, rather than agreement about what the best system should accomplish.

No ranking methods are value-free. Resume-based systems emphasize real-world wins and losses. Power ratings estimate the relative strength of all teams. A selection committee adds human judgment and context. A hybrid system combines and weights multiple forms of evidence. Each approach reflects a different philosophy of competitive merit and creates tradeoffs among predictive performance, deservingness, transparency, simplicity, fairness, and public trust.

Consequently, the central question is not simply, "Which ranking is correct?" A more useful starting point is "What is this ranking supposed to mean and what decision is it intended to support?" A system designed to predict future games may not be appropriate for determining which teams have earned championship access. Similarly, a method suitable for selecting a tournament field may not be the best for seeding it.\footnote{A post on an NCAA webpage clarifies that the basketball selection committee tends to use results-based (win/loss) methods more than predictive (power rating) methods for selection; still, they prefer predictive methods for seeding  \citep{McGuire2025,ESPNSportsAnalytics2016}.}

This paper provides a framework for discussing those distinctions. First, it separates prediction, retrospective evaluation, selection, and seeding as distinct decision tasks that require different standards of success. Second, it organizes twenty-five mathematical, institutional, and competitive qualities into a common framework for evaluating ranking systems. Third, it compares major families of ranking methods according to the philosophies of merit they embody rather than treating their numerical outputs as directly interchangeable. Fourth, it uses nine NCAA Division I sport settings to show why differences in schedule density, scoring, and competitive-network connectivity make universal multi-sport assumptions problematic.

The paper is not intended to defend a particular ranking system or identify one universally superior method. It is conceptual and descriptive: it proposes a framework and testable research agenda, but it does not claim that one method has been proved superior. The objective is to provide coaches, administrators, analysts, and governing bodies with a common vocabulary for evaluating whether a ranking system is appropriate for its intended purpose. 

\section{Ranking tasks: Prediction, retrospective evaluation, selection, and seeding}
Before debating formulas, stakeholders should identify the decision a ranking is intended to support. A method can simultaneously be excellent for forecasting and a poor basis for championship selection; conversely, a resume-based selector can accurately reflect past accomplishment while offering little information about future performance. Separating tasks is one way to clarify necessary tradeoffs.\footnote{The computer is "only the little black box" that outputs rankings according to the set of principles and tradeoffs devised by a human. Rankings are the quantitative result of the particular tasks a method is designed to solve \citep{Stern2006}.}

\subsection{Prediction}
Predictive systems estimate how teams are likely to perform in future games, and methods designed primarily for this purpose include power ratings, margin-based models, betting-market estimates, and, more recently, machine-learning models. 

The performance of a predictive system can be evaluated using future results, including how accurately they predict winners, scoring margins, or win probabilities.\footnote{According to \citet{Barrow2013}, among the most predictive ranking methods are least squares (regression-based) and random-walker methods, especially those "utilizing score-differential data."} Prediction therefore asks: Which team is most likely to perform better in the rest of the season or post-season tournament? 

This is an important question, but not necessarily the same as asking which team has earned selection based on its performance during the season. A team may be highly rated because it is, for some reason, expected to succeed in the future even though another similar team has compiled the more "deserving" body of work.

\subsection{Retrospective Evaluation}
Retrospective systems evaluate what teams have accomplished in games already played and commonly consider wins and losses, opponent quality, strength of schedule, head-to-head results, and/or other evidence from a team's complete body of work. Unlike predictive systems, their primary purpose is not to forecast future performance. 

Retrospective evaluation therefore asks: Which team has earned the stronger record of accomplishment? This question is closely associated with the idea of the "most deserving" team. 

Deservingness is not self-evident: different systems may emphasize total wins, schedule difficulty, head-to-head results, or other criteria and consequently reach different conclusions.\footnote{\citet{Coleman2005} posits that "maximizing retrodictive accuracy" (retrospective accuracy, in other words), "is equivalent to minimizing" cases where "a past game's winner is ranked behind its loser." This definition of retrospective accuracy aligns with the common sense and oft-repeated idea that "you have to value winning" \citep{IrmerOrbit}.}

\subsection{Selection}
Selection is about converting evidence about teams into decisions about which have earned access to the championship tournament bracket. This task often considers retrospective achievement, predictions about team strength, scheduling or geographic fairness, victory in conference championships, or some combination of these objectives.\footnote{Quotas like those previously used by Division II football or the Paris 2024 Olympics might be justifiable, depending on the sporting landscape \citep{NCAA2025Football,FIBA2024}.}

Selection therefore asks: Which teams should receive access to the championship tournament and according to what definition of merit? 

This is not merely a mathematical question. The governing organization must decide which evidence is relevant, how to balance competing considerations, and how to justify consequential decisions to teams that are included or excluded from the tournament.

\subsection{Seeding}
Seeding begins after the championship field has been selected and is intended to arrange teams within the tournament bracket in a way that reflects relative quality and promotes competitive balance. Seeding often considers predictions about team strength more than selection, which, as discussed, typically stresses deservingness \citep{McGuire2025}. Seeding decisions may be influenced by geography, travel, hosting requirements, conference rematches, and other operational constraints.\footnote{In addition, seeding may or may not adjust for sudden changes to team strength such as, for instance, injury, potentially improving upon solely results-based methods \citep{Gray2012}.}

Note that a method appropriate for choosing which teams belong in the field is not automatically the best method for ordering them. 

Seeding therefore asks: How should the selected teams be placed within the tournament? 
\subsection{On the best vs. most deserving debate}
Ranking disagreements frequently arise when the terms "best" and "most deserving" are treated as though they mean the same thing. Whereas the best team is generally understood to be most likely to perform well in future competition, the most deserving is the one judged to have earned recognition through its completed body of work. The concepts overlap but produce different conclusions. 

Some methods combine elements of both predictive and retrospective approaches in an attempt to reconcile the underlying philosophical choice about which objective should receive priority.

\subsection{Conclusions on ranking tasks}
In practice, each ranking system should be evaluated according to the task(s) it was built to solve. 

Accuracy in forecasting future results, assessing a team's body of work, making tournament selections, and producing tournament seeding are related but often subtly distinct standards. Stakeholders who fail to identify which standard to apply risk appearing to disagree on evidence despite actually asking different questions.

\section{Stakeholder objectives and desired attributes}
Ranking systems should not be judged solely by, as they often are, whether they produce a preferred ordering of teams. Just because a team is ranked higher or lower than expected does not mean that a method has failed to "accurately" judge their performance. Moreover, reversing the position of one ranking to satisfy an imaginary ideal risks undesirable outcomes elsewhere on that same ranking list. 

Thus, one should clearly define the objectives and priorities an ideal ranking system should satisfy and evaluate the said system based on these qualities, instead of an imagined ideal list \citep{Keeney1993}. 

The following four clusters contain a "menu" of attributes that ranking systems may be designed to satisfy, providing a common vocabulary and a useful checklist for discussing ranking methods without presuming that any particular philosophy must prevail.

\begin{table}[ht]
\caption{Twenty-Five Attributes of Ranking Systems}
\label{tab:attributes}
\begin{tabularx}{\columnwidth}{@{}L@{}}
\toprule
\textbf{Cluster 1: Trust, openness, and accountability} \\
\addlinespace[2pt]
Transparency $\cdot$ Simplicity $\cdot$ Consistency $\cdot$ Accountability $\cdot$ Justifiability $\cdot$ Reproducibility $\cdot$ Institutional confidence \\
\midrule
\textbf{Cluster 2: Mathematical and analytical soundness} \\
\addlinespace[2pt]
Mathematical integrity $\cdot$ Data and inputs $\cdot$ Predictive validity $\cdot$ Stability throughout the season $\cdot$ Robustness $\cdot$ Task-specific accuracy \\
\midrule
\textbf{Cluster 3: Practicality and sustainability} \\
\addlinespace[2pt]
Ease of implementation $\cdot$ Administrative efficiency $\cdot$ Resistance to manipulation $\cdot$ Adaptability $\cdot$ Viability \\
\midrule
\textbf{Cluster 4: Fairness and competitive incentives} \\
\addlinespace[2pt]
Fairness $\cdot$ Scheduling $\cdot$ Respecting head-to-head results $\cdot$ Balancing the margin of victory $\cdot$ Incentive alignment $\cdot$ Geographic equity $\cdot$ Philosophical coherence \\
\bottomrule
    \end{tabularx}
\end{table}

\subsection{Sources of our attributes}
Attributes were assembled from several complementary sources: discussions with coaches and administrators on the MD1 Lacrosse Selection Criteria and Ranking Committee (SCR), published work on ranking and decision methods, public expectations about fairness and transparency, experience from years of operating sports rating systems, and practical lessons from existing NCAA approaches. 

They are intentionally broad and concern mathematical performance, governance, communication, incentives, and trust and are not assumed to be equally important in every sport or institutional setting. 

A useful discussion of these attributes should first identify non-negotiable(s), establish priorities among remaining attributes, and acknowledge that necessary trade-offs between priorities reflect the values and competitive structure of the sport in question.

\subsection{Cluster 1: Trust, openness, and accountability}

\subsubsection{Transparency}
Can stakeholders access the system's logic and inspect its formulas, weights, and inputs? If not, they cannot assess whether it is fair. Transparency is the foundation of public trust.\footnote{Both committee-based and complex mathematical models can suffer from a lack of transparency. One class of models beginning to be used for sports ranking are machine learning-based methods, which typically suffer from an inherent lack of interpretability \cite{Doshi2017}.}

\subsubsection{Simplicity}
Can coaches, administrators, media, and fans easily identify inputs and explain how the method works? Complexity is not a virtue. The simpler a system is, the more likely it is to be used correctly, discussed accurately, and accepted widely.

\subsubsection{Consistency}
Are the same rules applied across the teams and seasons without "ad hoc" changes? Inconsistent rules produce suspicions of unfairness. If two teams with identical bodies of work receive different treatment, the system loses credibility.

\subsubsection{Accountability}
Is a clearly identified person or institution responsible for explaining, defending, and answering for the rankings produced by their chosen method? By identifying who or what institution is accountable for the results of a ranking system, stakeholders ensure against arbitrary decisions.

\subsubsection{Justifiability}
Does the explanation for ranking one team above another correspond to what actually happened on the field? A system can be transparent but misguided and the reasons given for a ranking should be grounded in real competitive outcomes. 

\subsubsection{Reproducibility} 
Can an independent, qualified analyst reproduce the same results given the same data and exact methodology? Reproducibility is a safeguard against hidden manipulation or mistakes and opens the door for external validation, including "what-if" analysis by coaches and fans.

\subsubsection{Institutional confidence}
Do stakeholders trust the institution overseeing the rankings to act fairly, competently, and without improper influence? Even mathematically sound systems can lose credibility if the body overseeing them is perceived as partial, inconsistent, or insufficiently accountable. 

\subsection{Cluster 2: Mathematical and analytical soundness}
\subsubsection{Mathematical integrity}
Are the system's operations, assumptions, and relationships mathematically valid? Mathematically sound methods use well-defined procedures and valid reasoning to produce results that are reasonably stable.

\subsubsection{Data and inputs}
Does the system use all relevant information and justify any meaningful data it excludes? A system that ignores legitimate data may produce rankings that feel incomplete, biased, or arbitrary. 

\subsubsection{Predictive validity}
Does the system predict future performance better than chance or other reasonable benchmarks? Predictive validity is one potential test for whether a ranking succeeds in capturing team strength.\footnote{\citet{Barrow2013} tests predictive validity in a variety of goal-differential and win-loss metrics, finding that the least squares methods based on the margin of victory are among the most predictive.} 

\subsubsection{Stability throughout the season}
Do rankings stabilize as additional games are played? Early-season rankings are based on limited evidence and may change substantially as schedules become more connected, so a credible system should generally become more stable as results accumulate, reaching a stable state near the end of a season.\footnote{\citet{Ochieng2022} provides an approach for comparing rating methods based on stability over time using rolling and expanding windows, finding that Massey and PageRank methods are more stable than win percentage, least square, and eigenvector methods.} 

\subsubsection{Robustness}
Does the ranking system remain reasonably stable when small changes are made to the data or assumptions? Large changes suggest overfitting and measurement error and reduce confidence in the method's rankings. 

\subsubsection{Task-specific accuracy}
Does the system accurately perform its ranking task (prediction, retrospective evaluation, selection, or seeding)? Accuracy cannot be evaluated without knowing what success looks like.  Strong performance in one task does not imply accuracy in another.\footnote{Whereas predictive accuracy measures future outcomes, retrospective accuracy might require other standards.}

\subsection{Cluster 3: Practicality and sustainability}
\subsubsection{Ease of implementation}
Can the system be implemented without specialized expertise or expensive infrastructure? Ease of implementation goes hand-in-hand with simplicity and verifiability.

\subsubsection{Administrative efficiency}
Can rankings be produced accurately and on time for committees, teams, media, and fans? Slow approaches are assumed to be worse than equivalent quicker ones.

\subsubsection{Resistance to manipulation}
Does the system resist strategies of untoward manipulation such as strategic scheduling, running up the score, or other gamesmanship? Procedure should not influence team strategy or behavior before and during a contest.

\subsubsection{Adaptability}
Can the system accommodate changes in sport structure, scheduling, and data availability without losing its governing principles? The evolution of a sport, realignment of its conference structure, or slight adjustment in data sources should not require a ground-up rebuild of the ranking system's methodology.

\subsubsection{Long-term viability}
Will the system remain effective, relevant, and well-accepted across many seasons? Trusted systems produce reliable results despite changing conditions.

\subsection{Cluster 4: Fairness and competitive incentives}
\subsubsection{Fairness}
Do teams with similar competitive achievements receive similar treatment in the rankings? Fairness is not purely mathematical; it depends on shared intuitions about what deserves recognition.

\subsubsection{Schedule integrity}
Does the system properly account for schedule difficulty and reward performance rather than mere exposure to strong opponents? Schedule difficulty provides important context for a team's record, however, merely facing strong opponents (thanks to conference affiliation or geographic boundaries) is not as important as performing successfully against them. 

A credible system distinguishes strength of opposition from accomplishment against that opposition.\footnote{Evidence of conference bias in selection and seeding like that detailed in \citet{Coleman2010} and \citet{Paul2015} might relate to a failure to accurately account for scheduling differences.}

\subsubsection{Respect for head-to-head results}
When two teams are otherwise comparable, does the system appropriately reflect the result of their head-to-head meeting? In competitive sports, head-to-head results are regularly seen as the "gold standard" by coaches, teams, and the public.\footnote{\citet{Coleman2005} points out that, while results should be maximally respected, situations often arise in which "team A beats team B, team B defeats team C, and team C beats team A," or a more complicated cycle of more than 3 teams.}

\subsubsection{Balancing margin of victory data}
If making use of margins of victory, does the system extract meaningful information from scores without introducing noise from "garbage time" or encouraging teams to run up the score? Margins of victory can distinguish between narrow wins and dominant performances, but they must be handled carefully.\footnote{"A 15-run win isn't five times more informative than a 3-run win," as noted by \citet{Alexander2026}, but especially at low sample sizes, margin of victory can make win percentage predictions "substantially better" \citep{Skeptical2011}.}

\subsubsection{Incentive alignment}
Does the system encourage behavior that makes the sport more competitive, fair, and sportsmanlike?\footnote{\citet{Goodhart1975}'s famous adage that "When a measure becomes a target, it ceases to be a good measure."}

\subsubsection{Geographic equity}
Does the system avoid systematically advantaging or disadvantaging teams because of geography, travel, or scheduling access? Geographic location affects travel costs and scheduling options.

\subsubsection{Philosophical coherence}
Does the system express a clear and defensible philosophy of competitive merit? A system without philosophical coherence produces rankings that may be mathematically correct but intuitively unsatisfying.

\subsection{Conclusion on attributes}

Together, these twenty-five attributes provide a broad vocabulary for discussing ranking systems. No system will score perfectly on every attribute. They are not a scoring formula and need not be weighted. Their purpose is to make visible and discussable the assumptions behind ranking decisions. 

\captionsetup[table]{justification=centering}

\begin{table*}[!b]
\centering
\caption{Selected Characteristics of Nine NCAA Division I Team-Sport Settings}
\label{tab:sport-characteristics}

\small
\setlength{\tabcolsep}{3pt}
\renewcommand{\arraystretch}{1.08}

\begin{tabular*}{\textwidth}
{@{\extracolsep{\fill}}lrrrrrcc@{}}
\toprule
\textbf{Sport}
& \textbf{Teams}
& \shortstack[r]{\textbf{Games}\\\textbf{per team}}
& \shortstack[r]{\textbf{Avg. total}\\\textbf{score}}
& \shortstack[r]{\textbf{Avg.}\\\textbf{margin}}
& \shortstack[r]{\textbf{Cross-conf.}\\\textbf{play}}
& \shortstack{\textbf{Head-to-head}\\\textbf{connectivity}}
& \shortstack{\textbf{Common-opponent}\\\textbf{connectivity}} \\
\midrule
Football (FBS) & 136 & 12--13 &  53 &     19 &     40\% & Low    & Medium \\
Basketball (M) & 365 & 31--32 & 150 & 12--16 & 35--55\% & Low    & Medium \\
Basketball (W) & 363 &     30 & 130 & 16--17 & 35--45\% & Low    & Medium \\
Baseball       & 308 & 55--60 &  13 &      5 & 45--60\% & Low    & Medium \\
Softball       & 308 & 50--60 &  10 &      5 & 50--70\% & Low    & High   \\
Lacrosse (M)   &  77 & 15--16 &  23 &      6 &     50\% & Low    & High   \\
Lacrosse (W)   & 133 &     17 &  23 &      7 &     50\% & Low    & High   \\
Ice hockey (M) &  63 & 35--40 &   6 &      2 &     30\% & Medium & High   \\
Ice hockey (W) &  45 & 35--45 &   5 &      3 &     25\% & Medium & High   \\
\bottomrule
\end{tabular*}

\par\vspace{4pt}
\begin{minipage}{0.96\linewidth}
\footnotesize
\textit{Note.} Figures are approximate and describe the
2025--26 academic year. Total score is the sum of both teams'
scores; margin is the absolute difference. Games per team
indicates approximate schedule length rather than a calculated
season average. Head-to-head and common-opponent connectivity
are qualitative estimates of the percentages of distinct team
pairs within each sport that met directly or shared at least
one opponent, respectively.
These categories illustrate broad scheduling differences;
they are not verified percentage calculations or measures
of pairwise comparisons resolved.
\end{minipage}
\end{table*}

\section{Structural challenges in intercollegiate competition}
Ranking systems built on sound principles still operate under difficult mathematical and institutional constraints. Teams play uneven schedules, data is limited, results are sensitive, and definitions of fairness, accuracy, and deservingness are contested. In addition, systems must satisfy the practical (and sometimes varied) needs of coaches, administrators, governing bodies, and the public. 

This section examines the principal challenges that confront all ranking systems and explains why no one method can completely eliminate uncertainty and controversy.

\subsection{Sparse data}
Intercollegiate schedules provide only a limited sample of the possible comparisons among teams. 

Most teams play only a small proportion of the other teams in their division, and the amount of cross-conference competition varies substantially by sport. Consequently, ranking systems must estimate relationships between teams that have not played one another and may have few common opponents. 

These limitations are most apparent in the large middle-tier conferences and among bubble teams, where small differences can change ranking lists, sometimes drastically. Final scores and scoring margins can provide additional information for ranking purposes.

\subsection{Scope and transferability}
Whereas the NCAA aims to apply a common algorithm for team strength ratings across NCAA Division II and III team sports in the NCAA Power Index (NPI), transferability should not be assumed merely because similar inputs are available. Sports and divisions differ in ways that influence how reliably a ranking system can compare teams. 

The data in Table 2 illustrate how the number of teams, games per team, schedule density, scoring frequency and variance, divisional and regional structure, and degree of cross-conference play can influence the network connectivity of sporting data and comparison paths among team pairs. 

More cross-conference play allows information to diffuse between teams; less play leaves conferences as relatively isolated clusters. As illustrated, the availability of head-to-head and common-opponent comparisons varies substantially across the nine settings. A common multi-sport algorithm may be administratively attractive, but it should be validated separately in each sport and division.

\subsection{Human bias}
Human judgment can introduce conscious or unconscious bias into ranking and selection decisions. 

Committee members may represent particular conferences, regions, or institutions, creating actual or perceived conflicts of interest. Mathematical systems such as the RPI or NPI can reduce reliance on human judgment, but their design still reflects human choices about data, weights, parameters, and definitions of merit. Neither committees nor algorithms eliminate bias entirely; credible governance should instead make potential sources of bias visible and establish procedures for limiting their influence. 

\subsection{Compromising on simplicity with "ad hoc" \\adjustments}
While simplicity is an important attribute for coaches, administrators, media, and the public's understanding of a method, additional components may be necessary to improve analytical performance, address unusual circumstances, or promote fairness. 

Each modification produces both benefits and costs, and must be weighed against its loss of transparency, reproducibility, or explainability. The central question is whether the incremental gain in, say, accuracy justifies the added complexity and potentially making the method harder to understand and defend.

\subsection{Iterative procedures}
Some ranking methods, like the NPI, require an iterative process to reach "convergence" in which teams' ratings depend partly on the ratings of their opponents. Because those opponent rankings are themselves being calculated, the system begins with initial values and repeatedly updates the rankings until subsequent changes become sufficiently small — hence, convergence. 

Iteration is not inherently a mathematical weakness, nor does it necessarily mean that a method is nonlinear. The primary concern is transparency: independent users must know the starting conditions, formulas, adjustment rules, stopping criteria, and treatment of exceptional cases to reproduce the published results. Without that information — or the original code — verification becomes considerably more difficult.

\subsection{Clustering and false precision}
Ranking systems often produce clusters of rankings whose estimated differences in team strength are small relative to the uncertainty of the underlying data. A formula may assign each team a distinct numerical value (even with multiple decimal places of precision) that nevertheless fails to correspond with meaningful competitive separation. 

This concern is especially important near selection and seeding boundaries, i.e. "on the bubble," where a small rating difference can produce a consequential institutional decision.

Systems should therefore report or examine sensitivity to perturbation, uncertainty, and the stability of close comparisons rather than assuming that every numerical ordering represents a clear difference in team quality or accomplishment.

\subsection{Gaming the system}
As mentioned earlier, teams may adjust their scheduling or game strategy to improve their rankings, a form of strategic response to adverse incentives. 

Win-loss algorithms may, for instance, incentivize programs to avoid risky match-ups and optimize strength of schedule calculations. Margin-informed systems may reduce some of these incentives by rewarding quality of performance at the potential cost of creating pressure to score even after a game's outcome is no longer in doubt. Methods may employ caps, diminishing returns, or game-state adjustments to discourage such behavior at the cost of introducing additional parameters. 

No ranking system is completely resistant to manipulation; the objective is to identify what incentives exist and ensure that they remain consistent with competitive integrity and sportsmanship.

\subsection{Choosing a goal cutoff}
Margin of victory-informed systems cap or taper scoring differentials to reduce the influence of extreme results and discourage unnecessary late-game scoring. 

A cap set too low may discard useful information about team strength, causing the system to approximate a binary win-loss model. A cap set too high may add little predictive information while increasing the risk of incentivizing bad sportsmanship or false prediction.

The appropriate threshold may differ by sport because scoring frequency, game structure, competitive balance, and expectations regarding sportsmanship vary. Caps should therefore be treated as parameters requiring empirical validation and review rather than as self-evident or universal.

\subsection{Valuing head-to-head results}
In the author's experience, a common objection to rankings reads something like: "We beat that team, so why are they ranked ahead of us?" Head-to-head results carry a unique intuitive and competitive importance. 

When all other factors are assumed to be equal, a victory by team A over team B should guarantee that team A earns a higher ranking, but one game may not outweigh evidence from an entire season, particularly if the teams have different schedules and if other results strongly favor team B that lost in a direct meeting or, notably, if team A lost to a team C which team B beat. 

A credible system should specify in advance how it handles head-to-head evidence and should be interpretable enough to explain when broader season performance overrides a head-to-head result.

\subsection{Disproportionally weak teams}
Conferences with especially weak opponents present a difficult design problem. Sometimes even a victory against a weak team can cause a team's rating to decline as a result of a given ranking system's strength of schedule calculation.

Game scores help avoid some of this effect by considering how a team performed, but they may over-reward extreme scoring differentials against weak teams unless margins are duly tapered or capped. 

A credible method should neither reward teams for scheduling weak opposition nor penalize them for playing required opponents, and coaches should know in advance how such games will affect their rankings. 

\subsection{Late-season games}
Games played later in a season may better reflect a team's strength come time for the championship tournament for various reasons, including roster changes, improvements in strategy, or improvements in physical fitness. 

Ranking systems must determine whether to weight games equally or to give greater weight to late-season games. Recency weighting may better reflect strength at the cost of undervaluing early victories and creating scheduling inequities in which teams cannot control when they play their most important games.

\subsection{Conclusion on structural challenges}
To recap: as mentioned, ranking systems for college sports face unique, sometimes contradictory challenges, including: sparse data with little transferability and connectivity, the risk of human bias, compromises between simplicity and ad-hoc improvements, the risk of non-convergence and false prediction, sportsmanship concerns, the salience of head-to-head results, the choice of a goal differential cutoff or recency bias, and, occasionally, conferences with especially weak teams. A perfect universally applicable system probably does not exist.

\section{Families of ranking methods}
Ranking systems can be grouped by structure and philosophy of merit. The below categories are not mutually-exclusive and real systems often combine elements from more than one. 

\begin{table}[tb]
\centering
\caption{Types of Ranking Systems and Examples}
\label{tab:ranking-types}

\small
\setlength{\tabcolsep}{4pt}
\renewcommand{\arraystretch}{1.12}

\begin{tabularx}{\linewidth}{
    @{}>{\raggedright\arraybackslash}X
    >{\raggedright\arraybackslash}p{0.25\linewidth}@{}
}
\toprule
\textbf{Type} & \textbf{Example} \\
\midrule
Win--loss resume              & RPI \\
Margin of victory                & Power ratings \\
Adjustable parameter (weighted)  & NPI \\
Expert judgment (committee)      & CFP \\
Multi-factor composite           & NET \\
Consensus                        & BCS \\
Market-based                     & Oddsmakers \\
Artificial intelligence or machine learning & --- \\
\bottomrule
\end{tabularx}

\end{table}

\subsection{Win-loss resume systems (ex: the Rating Percentage Index or RPI)}
These rating systems evaluate accomplishment through wins, losses, and opponent records. The Rating Percentage Index (RPI) is a familiar example of a win-loss resume method.

Win-loss resume systems are comparatively easy to describe and use inputs that are widely available across sports. They tend to include adjustments for whether teams won and the quality of their schedules to adjust for differing regular season schedules.

Their principal limitation is that they necessarily treat narrow and dominant victories as equally valuable: a virtue when data are plentiful, less-so when data are sparse. In college sports, win-loss systems may occasionally fail to consistently produce accurate or statistically significant solutions due to a lack of data connecting disparate teams or conferences.

\subsection{Margin of victory (MOV) systems (ex: Power rating systems)}
Margin of victory systems treat the size of a game result as valuable information about relative team strengths, usually employing safeguards like cutoffs to limit the influence of extreme scores and discourage unnecessary bad sportsmanship.

Margin-informed methods preserve information that win-loss systems discard, which has been shown to improve predictive accuracy in \citet{Barrow2013} and several other studies.

Their suitability depends on both analytical constraints (the data environment)\footnote{Margin of victory systems are often less necessary in conferences where each team plays every other.} and whether the incentives they generate are consistent with the sport's competitive values.

\subsection{Adjustable parameter (Weighted) Systems (ex: the NCAA Power Index or NPI)}
These systems combine multiple components through specific weights, modifiers, or other adjustable parameters, including (in the case of the NPI) win percentage, location weights, quality win bonuses, and a dial for the number of wins until unfavorable victories are dropped from consideration (i.e., minimum wins retained). 

The parameters or weights can be adjusted by ranking authorities to produce different outcomes suitable for specific sports, conferences, or seasons, allowing adaptation to differences in the length of the schedule, scoring patterns, the size of the home-field advantage, and the data environment. 

However, flexibility creates a justification problem, as stakeholders seek (especially when given adverse results) information about how each parameter was selected, what evidence supports it, and how sensitive the final rankings are to alternative values. Arbitrary parameters, even when established in advance and clearly documented, can be perceived as an attempt to produce a preferred ranking outcome.

\subsection{Expert judgment or committee-based systems (ex: the College Football Playoff Committee)}
This type of ranking system relies on a committee or panel to evaluate quantitative evidence alongside contextual information like injuries, unusual scheduling circumstances, late-season changes in strength, head-to-head context, and any other information that formal models may omit.

Challenges include transparency, reproducibility, consistency, and accountability. Outside observers often lack knowledge about how individual criteria were weighted, whether comparable cases were treated similarly, and what evidence proved to be ultimately decisive. 

Their legitimacy therefore depends on clearly stated criteria, procedures for managing conflicts of interest, and detailed explanations of consequential decisions.

\subsection{Multi-factor composite systems (ex: The NCAA Evaluation Tool, or NET, for basketball)}
These systems combine multiple types of information to represent several dimensions of team performance. NET, used in DI basketball, illustrates the broad approach, including 1) a metric for adjusted net efficiency based on points per possession, opponent points per possession, opponents strength, and the location of the game played and 2) a team value index based on wins and losses.

A multi-factor system may capture information that a simpler measure omits, potentially improving the evaluation of team strength, deservingness, and schedule performance. 

Its principal challenge is explainability. As the number of inputs, adjustments, and interactions increases, so does the difficulty of understanding why one team is ranked above another. 

\subsection{Consensus systems (ex: the old Bowl Championship Series or BCS, at least in part)}
These systems combine multiple rankings derived from various types of models or expert polls to produce a final result that does not depend entirely on one method's results. 

A consensus system reduces sensitivity to any unusual or spurious assumptions made by any single component method, revealing areas of broad agreement. 

Their value depends on the quality and diversity of the systems being combined. If components use similar assumptions, share the same biases, or measure different concepts of merit, combining them may reinforce weaknesses or produce a result that isn't coherently interpretable. A credible consensus system must explain why each component was included, how components were weighted, and what the combined result means.

\subsection{Market-based systems (ex: the Las Vegas oddsmakers or online platforms)}
These systems use betting lines or other market information to estimate expected game outcomes and, therefore, relative team strength. They often inherently and rapidly incorporate new statistical analysis, injury and other team news, public expectations, and judgments of market participants. Their performance can be evaluated against subsequent results.\footnote{Nick Saban pointed to oddsmakers as source of information about team strength after Alabama was not selected in 2022: "When they told me that we would be favored against three out of the four teams that got in the playoff, I'm like, why aren't we in the playoffs?"\citep{Flaherty2022}}

Market prices are not intended to reward season-long accomplishment or determine championship deservingness. They may also be influenced by relative liquidity constraints, betting biases, risk-management practices, and the different amount of attention afforded to certain teams or sports. 

While market information can provide a valuable predictive benchmark, it's inherently opaque, difficult or impossible to reproduce, and, may be inappropriate as a stand-alone selection system.

\subsection{Artificial intelligence or machine-learning systems}
AI methods identify patterns and uncover complex relationships that traditional models may overlook, potentially improving predictive performance when sufficient high-quality training data are available. 

Unfortunately, they are often opaque and many fail to set independently-verifiable, easily-repeatable standards for the evaluation of team strength.

Their applicability to the four sports ranking tasks identified above requires further study.

\subsection{Randomization}
While not a system, randomization provides a useful boundary case. Results should be meaningfully distinguishable from random chance, not just at the top or bottom of a ranking list but also "on the bubble."

\section{Common myths about ranking systems}
Discussions about ranking systems are often shaped by assumptions that appear reasonable but do not withstand careful examination. These misconceptions obscure the actual purpose of rankings, exaggerate the capabilities of mathematical models, and confuse disagreements about value-choices with technical disputes.

Table 4 (top of next page) highlights a few myths and/or misconceptions worth keeping in mind when choosing an optimal system.

\begin{table}[H]
\centering
\normalfont
\caption{Common Myths About Sports Rankings}
\label{tab:ranking-myths}

\small
\renewcommand{\arraystretch}{1.05}

\begin{tabularx}{\linewidth}{
    @{}>{\raggedright\arraybackslash}X@{}
}
\toprule

\textbf{Accuracy is all that matters.}\newline
Transparency, fairness and trust are also important. \\
\addlinespace[6pt]

\textbf{The best team is automatically the most deserving.}\newline
Defining deservingness requires value judgments. \\
\addlinespace[6pt]

\textbf{Computer rankings are entirely objective.}\newline
Computers are machines that do arithmetic and logic. They omit bias only if humans design their algorithms to do so. \\
\addlinespace[6pt]

\textbf{More complexity equals better rankings.}\newline
Complexity can cause overfitting and confuse underlying assumptions. \\
\addlinespace[6pt]

\textbf{Artificial intelligence will eventually create
the perfect ranking system.}\newline
It may identify complex patterns and improve
predictions, but it cannot determine
what a ranking ought to value. \\
\addlinespace[6pt]

\textbf{Controversy means failure.}\newline
Disagreement is inevitable given incomplete schedules, conflicting definitions of merit,
and genuine uncertainty. \\

\bottomrule
\end{tabularx}
\end{table}

\section{Practical framework for choosing a system}
A productive evaluation should begin with a purpose rather than a preferred method, formula, or outcome. The following framework provides a sequence applicable for use by governing bodies, committees, coaches, and analysts.

\subsection{Determine the ranking task and define "merit"}
State whether this system is intended for prediction, retrospective evaluation, championship selection, seeding, or some combination of these ends.

Then specify what the ranking of one team over another is supposed to indicate about those two teams. Is the higher ranked team stronger, more deserving, more likely to succeed in a tournament, or some combination of these concepts?

\subsection{Establish priorities and nonnegotiables}
Decide which attributes, of those listed in section 3, this ranking system should satisfy. Options may include transparency, reproducibility, predictive validity, resistance to some type of gaming, or others. Distinguish priorities from requirements and acknowledge any inherent conflicts between them.

\subsection{Match data and methodology to task}
Determine which data should legitimately influence rankings. These data potentially include game results such as wins and losses, scoring margins, and opponent quality or other factors such as injury reports, market information, and human intuition. 

Build a ranking methodology using one or more of those identified in section 5 to effectively convert data into a ranking list.

\subsection{Test close cases and unintended consequences}
A ranking system should be evaluated on relatively difficult cases and not just on the best and worst teams in a sports division. Sensitivity analysis should examine whether small changes in results, weights, caps, or assumptions produce outsized changes in outcomes. Such testing helps distinguish meaningful differences from methodological instability.

Ensure that your ranking system's rules do not unduly influence team strategy or behavior by incentivizing, for instance, unfair scheduling or score manipulation.

\subsection{Specify governing bodies and communication procedures and conduct regular reviews}
Carefully document methodology, data sources, rules in case of tight decisions, and procedures for revising the system in the future. Assign responsibility for explaining and defending the rankings.

Regular reviews are an appropriate part of ranking, but changes should be evidence-based, transparent, and applied consistently, rather than just in response to a single controversial result.

\section{Conclusions}
The central lesson of this paper is straightforward: ranking intercollegiate teams involves selecting formulas, committee processes, or hybrid approaches that make choices along the lines of competing philosophies of merit.

Prediction, retrospective evaluation, championship selection, and seeding are related but distinct tasks. A system designed to forecast future performance may not also identify the teams with the strongest regular season resumes. A method appropriate for selecting a championship field may not be the best for seeding it.

The structure of intercollegiate competition limits what rankings can accomplish. Sparse and uneven schedules, limited cross-conference competition, differences between sports and divisions, and closely matched bubble teams create uncertainty that no one method can eliminate. Limitations are especially important near tournament cutoffs, where small methodological choices may determine which teams are selected to compete.

For these reasons, ranking systems should be judged by more than just their "accuracy." Mathematical soundness and empirical performance are as important as transparency, simplicity, reproducibility, fairness, the alignment of incentives, explanatory fidelity, and institutional confidence. Complex mathematical algorithms, human judgment, market information, and AI-informed approaches can contribute useful evidence, but none are automatically appropriate for official use. 

To conclude: no ranking system is likely to eliminate controversy or incorporate all desirable attributes. Credible systems should instead seek to define their purpose, make tradeoffs visible, apply principles consistently, and explain their decisions in terms that stakeholders can understand. The most useful question in sports ranking is therefore not "Which ranking is correct?" but "What should this ranking mean?"

\newpage
\section{Statements and Declarations}
\subsection{Data}
The descriptive data summarized in Table 2 were obtained from d1field.com. Some are loose estimates from other sources. They are intended to illustrate the differences inherent in different NCAA sporting environments. 
\subsection{Acknowledgments}
This overview of ranking systems for NCAA sports was inspired by conversations with the NCAA Men's Division I Lacrosse Selection Criteria and Ranking Committee (SCR) about the creation of a ranking system called Powerwise (PWR) that may see future use. Matthew Bomparola prepared the final manuscript by rewriting large sections of the paper and editing any remaining text. He did not use artificial intelligence.
\subsection{Use of artificial intelligence}
The author used ChatGPT as an editorial assistant during the preparation of the first draft manuscript. The tool assisted with organization, drafting, clarity, and presentation. That being said, the research questions, conceptual framework, selection of attributes, data, substantive judgments, interpretations, and conclusions were developed and approved by the human author. The author accepts full responsibility for the accuracy, originality, and integrity of the manuscript. 

\onecolumn
\setlength{\bibsep}{2pt plus 0.3ex}
\urlstyle{same}

\bibliographystyle{plainnat}
\bibliography{references}

\end{document}